\documentclass{article}
\usepackage{spconf,amsmath,epsfig}
\usepackage{graphicx}
\usepackage{amsmath}
\usepackage{amssymb}
\usepackage{booktabs}
\usepackage{multirow}
\usepackage[T1]{fontenc}
\usepackage{color}

\let\OLDthebibliography\thebibliography
\renewcommand\thebibliography[1]{
  \OLDthebibliography{#1}
  \setlength{\parskip}{0pt}
  \setlength{\itemsep}{0pt plus 0.3ex}
}

\begin{document}\sloppy

% Example definitions.
% --------------------
\def\x{{\mathbf x}}
\def\L{{\cal L}}

% Title.
% ------
\title{Image Protection for Robust Cropping Localization and Recovery}
%
% Single address.
% ---------------
\name{Qichao Ying$^{1\star}$, Hang Zhou$^{2\star}$, Xiaoxiao Hu$^{1}$, Zhenxing Qian$^{1\ast}$\thanks{$^{\star}$ equal contribution. $^{\ast}$ corresponding author. This work is supported by the National Natural Science Foundation of China under Grants U20B2051.} , Sheng Li$^{1}$ and Xinpeng Zhang$^{1}$}
\address{$^{1}$Fudan University, China,
$^{2}$Simon Fraser University, Canada.\\
\{qcying20, xxhu20, lisheng, zxqian, zhangxinpeng\}@fudan.edu.cn, zhouhang2991@gmail.com\\
}

%Address and e-mail should NOT be added in the submission paper. They should be present only in the camera ready paper. 

\maketitle

\definecolor{green}{rgb}{0, 0.5, 0}
\definecolor{orange}{rgb}{0.8, 0.6, 0.2}
\definecolor{orange2}{rgb}{1.0, 0.6, 0.2}
\definecolor{red}{rgb}{1.0, 0.0, 0.0}
\definecolor{teal}{rgb}{0.0, 0.4, 0.4}
\definecolor{purple}{rgb}{0.65,0,0.65}
\definecolor{saffron}{rgb}{0.95,0.75,0.2}
\definecolor{turquoise}{rgb}{0.0,0.5,0.5}
\definecolor{black}{rgb}{0.0, 0.0, 0.0}
\definecolor{gray}{rgb}{0.5, 0.5, 0.5}

\newcommand{\greenmarker}[1]{{\color{black}#1}}
\newcommand{\redmarker}[1]{{\color{black}#1}}

\begin{abstract}
% Image cropping is a cheap yet effective operation of maliciously altering image contents. 
Existing image cropping detection schemes ignore that recovering the cropped-out contents can unveil the purpose of the behaved cropping attack. This paper presents \textbf{CLR}-Net, a novel image protection scheme addressing the combined challenge of image \textbf{C}ropping \textbf{L}ocalization and \textbf{R}ecovery. We first protect the original image by introducing imperceptible perturbations. Then, typical image post-processing attacks are simulated to erode the protected image. On the recipient's side, we predict the cropping mask and recover the original image. Besides, we propose a novel \textbf{F}ine-\textbf{G}rained generative \textbf{JPEG} simulator (FG-JPEG) as well as a feature alignment network to improve the real-world robustness. 
% Experiments demonstrate the effectiveness of CLR-Net.
Comprehensive experiments prove that the quality of the recovered image and the accuracy of crop localization \redmarker{are both} satisfactory.
\end{abstract}
\begin{keywords}
image protection, image cropping localization, image recovery, robustness
\end{keywords}

% Main text
\section{Introduction}
\label{sec:intro}
% ------------------------------------------
% ------------move to supplement------------
% ------------------------------------------
% With the exploding amount of images transmitted through the Internet, anyone can take a picture anywhere anytime. Digital images have largely replaced conventional photographs from all walks of life. The rapid advancements in digital
% image processing has made it extremely easy with the presence of user-friendly image editing software, and the edited images can be shared with others in seconds with social networking services. 
% Although these advances have benefited people’s lives, digital images can hardly enjoy the credibility of their conventional counterparts.
% Image forgery is the behaviors that deliberately alter the interpretation of the original image by modifying parts of its contents.
% ------------------------------------------
% ------------end to supplement------------
% ------------------------------------------

\begin{figure}[!t]
	\centering
    \includegraphics[width=0.49\textwidth]{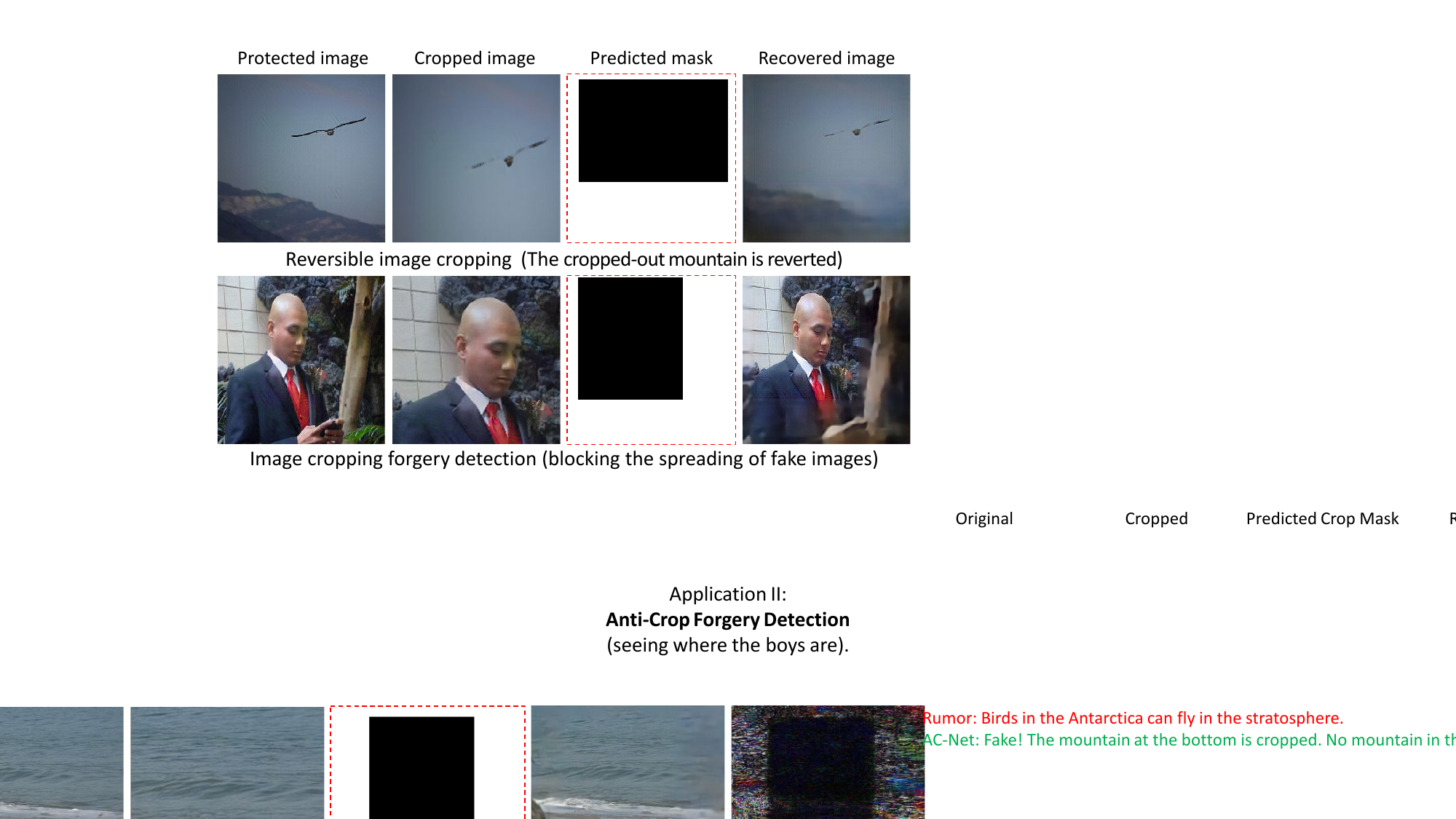}
	\caption{\textbf{Application of CLR-Net}. We transform images into protected ones via imperceptible signal injection. Once these protected images are cropped, CLR-Net can localize the cropped regions and recovers the cropped-out content.}
	\label{img_teaser}
\end{figure}

Image cropping is the process of removing unwanted areas of an image,
which creates focus and strengthens the composition.  
However, it can also be an extremely cheap and effective way to maliciously alter the underlying meaning. 
Fig.~\ref{img_teaser} shows two examples of malicious image cropping forgery. The attackers crop the images and mislead the audience with fake comments. 
Without the ground truth as a reference, people are easily deceived by what they observe from manipulated images. 
% What's worse, they might further circulate these fake images. As a result, fabricating stories by cropping can be a means for some politicians to influence public opinion.
% \begin{figure}[!t]
% \label{fig_teaser}
% 	\centering
% 	\includegraphics[width=0.48\textwidth]{figures/teaser_image.pdf}
% 	\caption{We transform images into the protected ones via imperceptible signal injection. Once these protected images are cropped, CLR-Net can localize the cropped regions and recovers the cropped-out content.}
% 	\label{img_teaser}
% \end{figure}
% Image cropping forgery detection is still an ill-posed problem
% % There are few previous arts for image crop forgery detection. 
% % The overwhelming majority focuses on cropping detection instead. 
% compared to traditional image manipulation detection.
% cropping forgery detection is much less investigated.
Existing image cropping detection schemes~\cite{fanfani2020vision,yerushalmy2011digital,li2009passive} only predict whether the image is cropped by searching for traces that expose image crops.
However, a successful cropping prediction or localization is not enough.
To fully investigate the intention of each suspicious cropping behavior, 
% as cropping-out different areas of an image can be with drastically varied intentions. 
% For example, how to distinguish benign cropping behaviors from those malicious attacks, such as discarding a visible watermark or removing a person? .
we need to locate the position of the crop in the original image plane, or even recover the original image.
In the literature, few schemes \cite{van2020dissecting} are developed for cropping localization and recovery.
% Moreover, these methods have certain limitations. 
% that make them applicable only to a subset of images. 
Moreover, \cite{van2020dissecting} require the images to be untouched and uncompressed so that tiny traces like chromatic aberration and vignetting exist. 
% In \cite{ying2021no}, a watermark for cropping localization must be shared with the recipient. 
% These constraints are hard to be satisfied in the real world. 
However, attackers can resize or compress the cropped image to counter that. 

\begin{figure*}[!t]
	\centering
	\includegraphics[width=1.0\textwidth]{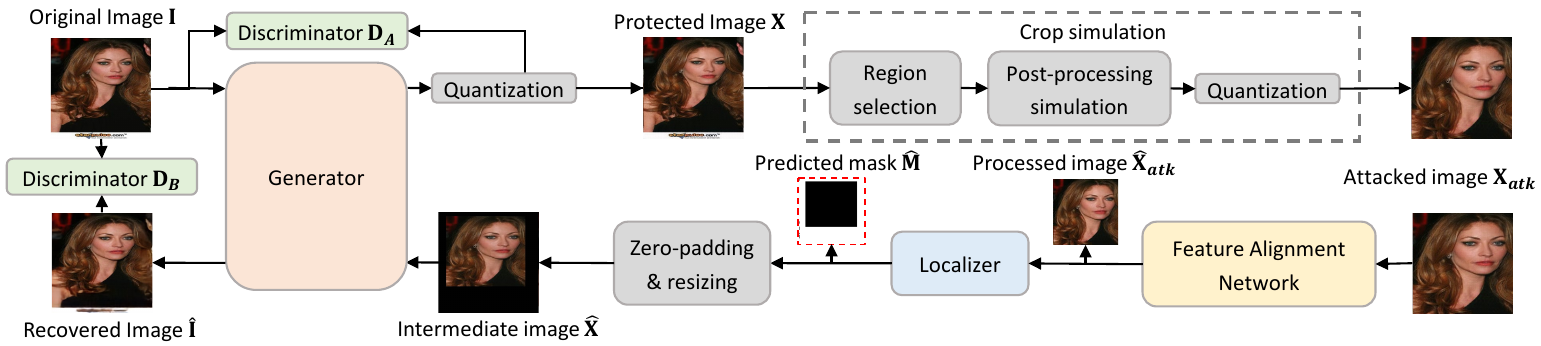}
	\caption{\textbf{Approach overview of CLR-Net.} The generator transforms the original image into the protected image. The attacked version is then aligned on the recipient's side, and the localizer predicts the cropping mask. We rectify the processed image by scaling and zero-padding. Finally, the inverse process of the generator \redmarker{estimates} the \redmarker{original} image.}
	\label{image_framework}
\end{figure*}

% We are motivated to develop a scheme that recovers the entire image from the cropped fragment, despite the presence of real-world image processing., e.g., JPEG compression, Gaussian blurring, etc. 
This paper explores the potential of image protection~\cite{zhu2018hidden,ying2021image} on image cropping localization and recovery.
We first protect the original image by introducing imperceptible perturbations. Then, typical image post-processing attacks are simulated to erode the protected image. On the recipient's side, we predict the cropping mask and recover the original image. 
% Besides, we propose a novel Fine-Grained generative JPEG simulator (FG-JPEG) as well as an feature alignment network to improve the real-world robustness of CLR-Net. 
% We propose a novel image protection network for Cropping Localization and Recovery (CLR-Net). When a targeted image is cropped, CLR-Net simultaneously localizes the cropped region and recovers the entire image. We propose a generator based on Invertible Neural Network (INN) \cite{xiao2020invertible} to transform an original image into its protected version. The difference before and after image protection is close to imperceptible. 
% Afterwards, we use an attacking layer to simulate malicious crop and benign image processing attacks. 
% On the recipient's side, a localizer is proposed to estimate the cropping mask and a confidence score. We rectify the image according to the cropping mask by resizing and zero-padding, and the inverse process of the generator recovers the original image. 
For enhanced robustness of CLR-Net in real-world applications, we innovatively propose a novel Fine-Grained generative JPEG simulator (FG-JPEG). 
Besides, a feature alignment network is proposed to minimize the performance gap against different lossy image operations. 
Comprehensive experiments demonstrate that CLR-Net accurately estimates the cropping mask and recovers the full image even if the attacker post-processes the cropped image. 
% The extensive ablation studies also prove the necessity of the network design as well as the training strategies.

It is well notifying that the task of CLR-Net is \textbf{Cropped Image Recovery} (CIR) that needs to exactly reconstruct what has been cropped-out by malicious users and predict the cropping mask as well.
The goal is different from outpainting that only extend the bordering of a given image using hallucination, i.e., providing non-unique solutions that are visually plausible.
CIR is important in many forensics or judicial scenarios that requires exact recovery to see what is cropped-out.
In this case, hallucinating possible outcomes using outpainting cannot be accepted.

% The main contributions of this paper are two-folded. 
% % \begin{itemize}
% % \item
% 1) We propose the first scheme for image cropping localization and recovery based on image protection.
% % \item
% % 2) Imperceptibility of the watermark embedding, high quality of the recovered images and high accuracy of cropping localization can be simultaneously ensured by CLR-Net.
% % \item
% 2) A feature alignment network, as well as a novel generative JPEG simulator named FG-JPEG, are proposed that noticeably improves the real-world robustness of CLR-Net. 
% \end{itemize}
% The robustness is strengthened with this design compared to applying previous differentiable attack simulators.
% \section{Related Works}
% \label{sec:related}
% \input{related_works}
\section{Method}
\label{section_method}
% ------------------------------------------
% ------------move to supplement------------
% ------------------------------------------
% \begin{figure}[!t]
% 	\centering
% 	\includegraphics[width=0.49\textwidth]{figures/fig2.pdf}
% 	\caption{\textbf{Problem statement of CLR-Net.} We decompose image cropping into several steps and build invertible functions for image cropping recovery. RS, IP, ZR respectively denotes \textit{region selection}, \textit{image post-processing} and \textit{zero-padding \& resizing}. Cropping localization is omitted for simplicity.}
% 	\label{image_problem_statement}
% \end{figure}
% ------------------------------------------
% ------------move end------------
% ------------------------------------------

% We first state the problem of image \redmarker{cropping} localization and recovery, and then introduce the network modeling accordingly. Training details and the objective loss functions of CLR-Net are also specified. Fig.~\ref{image_problem_statement} illustrates the problem statement of CLR-Net. The sketch of our modeling framework is presented in Fig.~\ref{image_framework}. 
% \subsection{Problem modeling}
% \label{section_problem}
% \greenmarker{\noindent\textbf{Formulation of cropping localization and recovery.}}
\subsection{Approach overview}
\noindent\textbf{Image Cropping Localization and Recovery (CLR).} 
Cropping and its recovery can be represented by Eq.~(\ref{equation_problem_statement}). 
\begin{equation}
\label{equation_problem_statement}
\begin{gathered}
\mathbf{I}_{\emph{atk}}=\emph{IP}\left(\emph{RS}\left(\mathbf{I},\mathbf{M}\right)\right), \\
\hat{\mathbf{I}}=f_{\emph{rec}}\left(\emph{ZR}\left(\mathbf{I}_{\emph{atk}},\hat{\mathbf{M}}\right)\right),
\hat{\mathbf{M}}=f_{\emph{M}}(\mathbf{I}_{\emph{atk}}),
\end{gathered}
\end{equation}
where $\mathbf{I}$ and $\mathbf{I}_{\emph{atk}}$ denote an image and its attacked version. 
$\mathbf{M}$ denotes the cropping mask. 
$\emph{RS}(\cdot)$ denotes \textit{region selection} that discards the information outside the selected region. 
$\emph{IP}(\cdot)$ denotes \textit{image post-processing} that truncates the image and post-processes the rest of the image using lossy operations, e.g., blurring.
Our task is respectively to \redmarker{estimate} $\mathbf{M}$ and $\mathbf{I}$ given $\mathbf{I}_{\emph{atk}}$.
% Usually, an additional image resizing is involved that transforms the attacked image into an arbitrary size. 
We employ two function\redmarker{s} $f_{M}(\cdot)$, $f_{\emph{rec}}(\cdot)$ respectively for CLR,
where $\hat{\mathbf{M}}$ and $\hat{\mathbf{I}}$ are respectively the \redmarker{estimated} cropping mask and the recovered image.
% , with $\mathbb{E}(\hat{\mathbf{I}})=\mathbf{X}$ and $\mathbb{E}(\hat{\mathbf{M}})=\mathbf{M}$.
$\emph{ZR}(\cdot)$ denotes \textit{zero-padding and resizing} that first zero-pads the attacked image according to the \redmarker{estimated} cropping mask and then resizes the padded image into the original size.
In most cases, few clues of the cropped-out contents can be found in the rest of the image.
So directly solving Eq.~(\ref{equation_problem_statement}) will lead to sub-optimal solutions where the destroyed contents are hallucinated rather than truthfully recovered.
% Note that the attacker can \redmarker{arbitrarily} reshape the image, and previous work~\cite{pfennig2012spectral} can only infer the original size approximately but not accurately. Therefore, we hypothesize that the original sizes of the images are fixed or specified by the recipient. 
% If the cropping localization is correct, $\emph{ZR}(\cdot)$ is the exact inversion of $\emph{RS}(\cdot)$. 

% \noindent\textbf{Introduction of image protection.}
% Modeling Eq.~(\ref{equation_problem_statement}) is ill-posed.} 
\noindent\textbf{Image protection for CLR. }
To address the above issue, we propose
robust image protection for CLR by
% , so that even if $\mathbf{X}$ is attacked with randomized $\emph{IP}(\cdot)$ and $\emph{RS}(\cdot)$, we can predict $\mathbf{M}$ and \redmarker{estimate the original} $\mathbf{I}$ with high fidelity.
employing a third function $f_{prt}(\cdot)$ to embed deep representations of an original image, denoted as $\mathbf{I}$, into itself.
Let
$\mathbf{X}=f_{\emph{prt}}(\mathbf{I})$
% To model the above functions, 
% with Eq.~(\ref{equation_problem_statement}), Eq.~(\ref{equation_problem_statement}) as well as $\mathbb{E}(f_{prt}(\mathbf{X}))=\mathbf{I}$ and $\mathbb{E}(f_{prt}(\hat{\mathbf{M}}))=\mathbf{M}$, we have
% % \mathbb{E}\left(f_{\emph{rec}}\left(\mathbf{X}_{\emph{atk}}\cdot\left(1-\hat{\mathbf{M}}\right)\right)\right)
% \begin{equation}
% \label{equation_invertible}
% \mathbb{E}\left(f_{\emph{rec}}\left(\emph{ZR}\left(\emph{cs}\left(\emph{RS}\left(f_{\emph{prt}}\left(\mathbf{I}\right),\mathbf{M}\right)\right),\hat{\mathbf{M}}\right)\right)\right)=\mathbf{I}.
% \end{equation}
% Accordingly, 
, and we reformulate Eq.~(\ref{equation_problem_statement}) as a pair of invertible image embedding and recovery functions, i.e., 
$\mathbf{X}=f_{\emph{prt}}\left(\mathbf{I}\right)$ and $\hat{\mathbf{I}}=f_{\emph{rec}}\left(\emph{ZR}\left(\mathbf{X}_{\emph{atk}},\hat{\mathbf{M}}\right)\right)$. 
We expect that $\mathbf{X}\approx\mathbf{I}$, $\hat{\mathbf{I}}\approx\mathbf{I}$ and $\hat{\mathbf{M}}\approx\mathbf{M}$, regardless of the lossy image operations.

The sketch of modeling image cropping and its localization \& recovery is presented in Fig.~\ref{image_framework}. 
We utilize an attack layer $\mathcal{A}$ to implement \textit{region selection} and \textit{image post-processing} in the real-world application.
The localizer $\mathcal{L}$ \redmarker{estimate}s the cropping mask $\hat{\mathbf{M}}$ and the confidence score $S$.
After zero-padding and resizing, we send $\hat{\mathbf{X}}$ into the generator $\mathbf{G}$ and inversely run the network to \redmarker{recover} $\mathbf{I}$.
We also introduce two discriminators $\mathcal{D}_{A}$, $\mathcal{D}_{B}$ to facilitate the imperceptibility of image protection and the quality of the recovered images.

\noindent\textbf{Mechanisms for enhanced robustness.}
First, though previous researches have proposed several step-by-step JPEG simulation methods~\cite{shin2017jpeg,zhu2018hidden}, the algorithm and the quantization tables in these schemes are immutable, which is in contrast with real-world compression adaptively controlled by the quality factor and the image content. 
% As a result, the neural networks can overfit and lack real-world robustness.
To address this, 
we tempt to add more flexibility in JPEG simulation by proposing the novel generative FG-JPEG simulator, as specified in Section~\ref{section_detail}.
% We propose a generative JPEG simulator called FG-JPEG to transform plain-text images into simulated JPEG images. 

% ------------------------------------------
% ------------move to supplement------------
% ------------------------------------------
% \greenmarker{The attacks include the following: (1) \textit{rescaling}: we resize the image by an arbitrarily resizing rate $r\in[50\%,150\%]$, (2) \textit{median blurring}, we blur the image using median filter whose kernel size $k$ is arbitrary selected from ${3,5}$, (3) \textit{Additive White Gaussian Noise} (AWGN), which adds Gaussian noise evenly on the image, where the standard value s ranges from zero to one. (4) Gaussian blurring, which is similar to the median blurring but the kernel is different.  
% (5) \textit{JPEG compression.} We exceptionally apply our own FG-JPEG to produce high-fidelity JPEG images with controllable quality factor. The design of FG-JPEG is specified in Section~\ref{section_FGJPEG}. The considered quality factor ranges from $[50,100]$.}
% Afterwards, $\mathbf{X}_{\emph{atk}}$ is again quantized into to the 8-bit integer format.
% ------------------------------------------
% ------------move end------------
% ------------------------------------------

% ------------------------------------------
% ------------move to supplement------------
% ------------------------------------------
\begin{figure}
	\centering
	\includegraphics[width=0.49\textwidth]{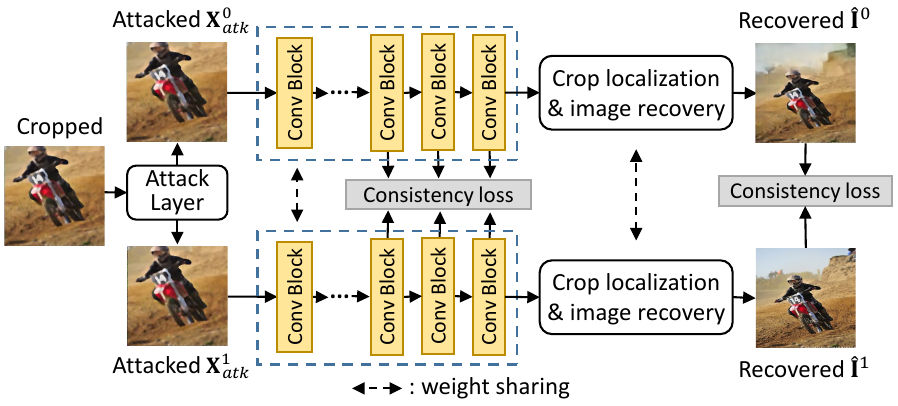}
	\caption{\textbf{The Siamese architecture of the feature alignment network.} The attack layer generates two attacked views of the cropped image using \redmarker{arbitrarily}-sampled attacks. We minimize the consistency loss between the image features as well as the performance gap under the two attacks.}
	\label{preprocessor}
\end{figure}

% ------------------------------------------
% ------------move to supplement------------
% ------------------------------------------
\begin{figure}[!t]
	\centering
	\includegraphics[width=0.48\textwidth]{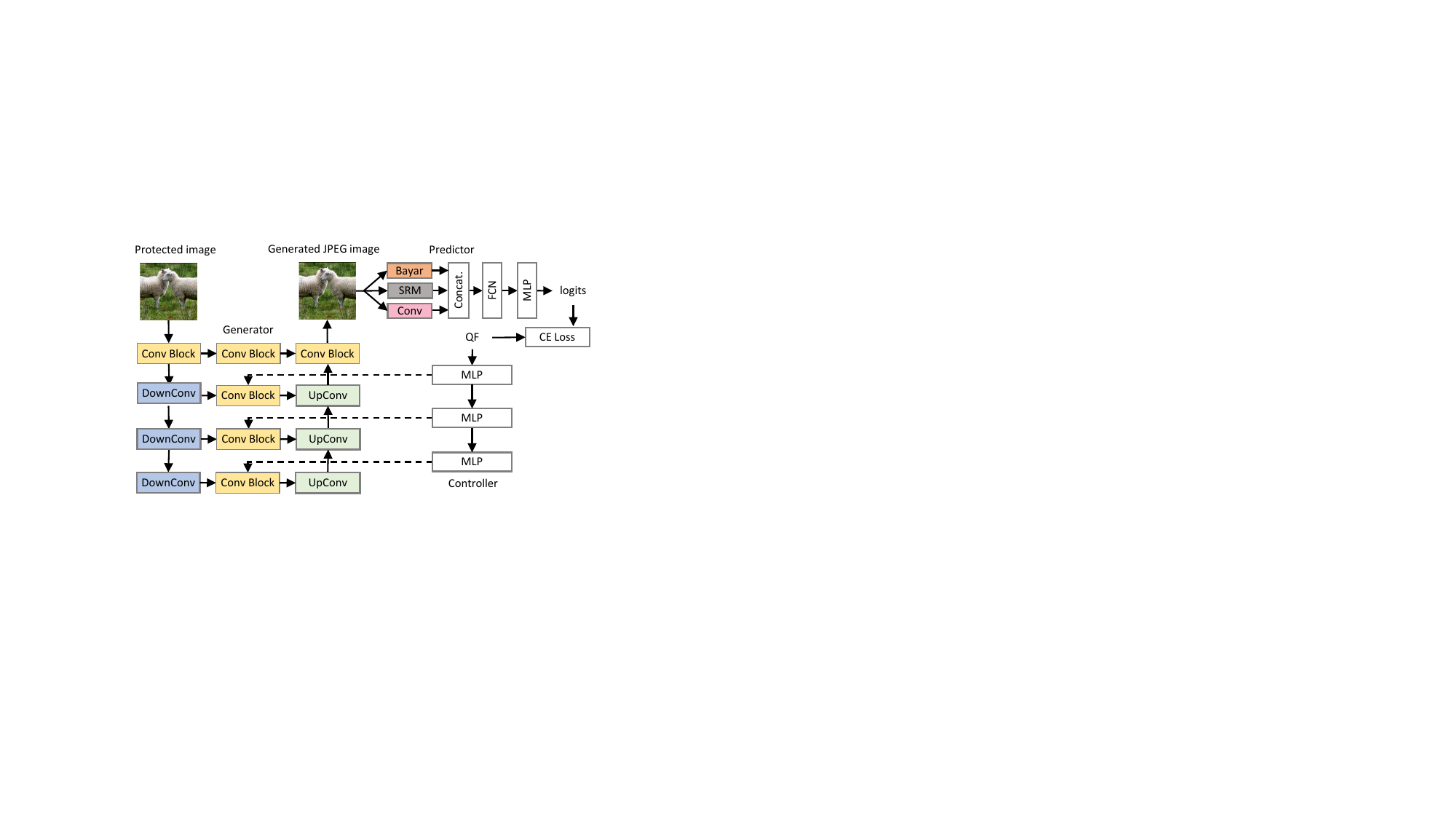}
	\caption{\textbf{Network architecture of FG-JPEG.} The generator takes the protected image and simulates the JPEG compression, which is conditioned by the controller. The predictor classifies the QF of the generated images.}
	\label{image_conponents}
\end{figure}
% ------------------------------------------
% ------------move end------------
% ------------------------------------------
% \noindent\textbf{feature alignment network. }
% , which maps the arbitrarily attacked images into a more clustered distribution.
% We align the received attacked image $\mathbf{X}_{\emph{atk}}$ using a distillation-based pairwise network $\mathbf{P}$, which is ahead of crop localization and recovery.
% \greenmarker{Given attacked versions of an $\mathbf{X}$, we use a Siamese network to transform these copies by minimizing their mutual differences. The purpose is to} help the subsequent network identify and utilize the features most tolerant to all possible digital image post-processing attacks. 
Second, $\emph{IP}(\cdot)$ is likely to contain a large solution space in the real world. 
% Different image post-processing attacks share varied characteristics, such as randomized addition or removal of higher-band information. 
Using the attack layer can only provide multi-task guidance, which does not explicitly circumvent uneven network performance against different kinds and strengths of attacks.
We introduce a feature alignment network $\mathcal{F}$ to mitigate this issue, as sketched in Fig.~\ref{preprocessor}.
In each iteration, we let the attack layer arbitrarily select two kinds of image post-processing attacks and generate two corresponding images $\mathbf{X}_{\emph{atk}}^0$ and $\mathbf{X}_{\emph{atk}}^1$ from a same $\mathbf{X}$. 
On cropping $\mathbf{X}_{\emph{atk}}^0, \mathbf{X}_{\emph{atk}}^1$, we use a shared mask $\mathbf{M}$.
Thereafter, $\mathcal{F}$ transforms $\mathbf{X}_{\emph{atk}}^0, \mathbf{X}_{\emph{atk}}^1$ into $\hat{\mathbf{X}}_{\emph{atk}}^0, \hat{\mathbf{X}}_{\emph{atk}}^1$ using the shared parameters.
Afterward, we respectively conduct cropping localization and image recovery based on the two views, where the \redmarker{estimated} masks, as well as the recovered images, are respectively denoted as $\{\hat{\mathbf{M}}^0,\hat{\mathbf{I}}^0\}$, $\{\hat{\mathbf{M}}^1,\hat{\mathbf{I}}^1\}$. 
We restrict that the performance of CLR-Net against the two attacks should be close between each pair, i.e., $\hat{\mathbf{M}}^0\approx\hat{\mathbf{M}}^1$, $\hat{\mathbf{I}}^0\approx\hat{\mathbf{I}}^1$. 
We also minimize the distance between the representations $\phi_{n}^0$, $\phi_{n}^1$ extracted from $\hat{\mathbf{I}}^0$, $\hat{\mathbf{I}}^1$. Third, we prevent $\mathcal{F}$ from drastically changing the image, i.e., $\hat{\mathbf{I}}^0\approx\hat{\mathbf{I}}$. 

\begin{figure*}[!t]
	\centering
	\includegraphics[width=1.0\textwidth]{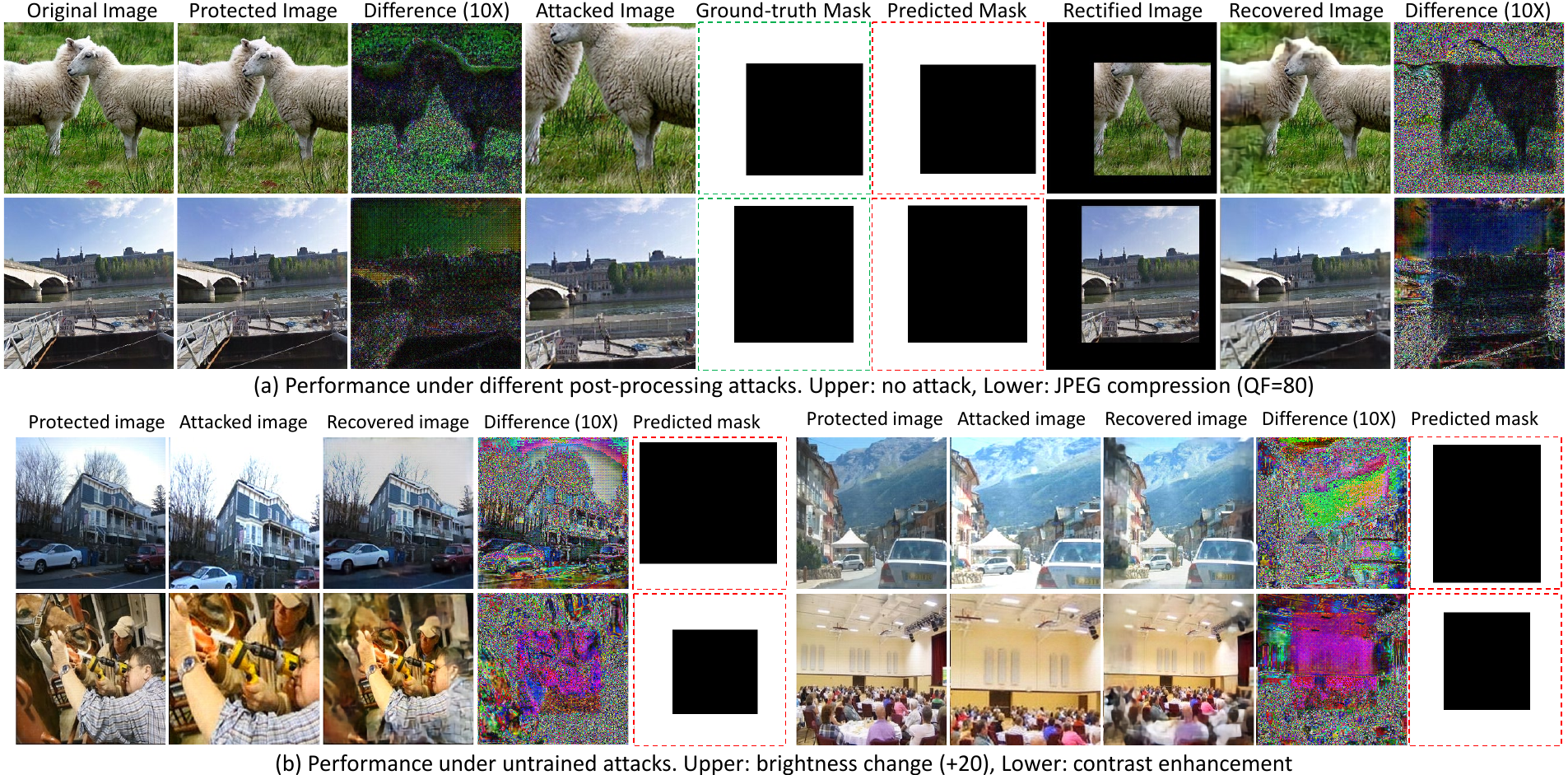}
	\caption{\textbf{Qualitative analysis of CLR-Net against different lossy image operations and varied survival rate.} The testing image are from COCO, ImageNet and Places.}
	\label{image_main_results}
\end{figure*}

\noindent\textbf{Objective functions.}
% \label{section_loss}
% The objective functions include the protection loss ${L}_{prt}$, the cropping localization loss ${L}_{\emph{loc}}$, the recovery loss ${L}_{\emph{rec}}$, the consistency loss ${L}_{\emph{cs}}$ and the adversarial losses ${L}_{\emph{adv}}^{\mathcal{D}_{A}}$ and ${L}_{\emph{adv}}^{\mathcal{D}_{B}}$. There is also the \greenmarker{JPEG} simulation loss ${L}_{\emph{jpg}}$ for FG-JPEG. In the below equations, $\alpha,\beta,\gamma,\theta,\eta$ are hyper-parameters.
% \greenmarker{\noindent\textbf{Generator.}}
% The protection loss ${L}_{prt}$ and the recovery loss ${L}_{\emph{rec}}$ respectively encourage $\mathbf{X}$ and $\hat{\mathbf{I}}$ to resemble $\mathbf{I}$. 
We use the popular $\ell_1$ loss term to measure the distance between images.
\begin{equation}
{L}_{\emph{prt}}=\lVert\mathbf{I}-\mathbf{X}\rVert_{1}, {L}_{\emph{rec}}=\lVert\mathbf{I}-\hat{\mathbf{I}}\rVert_{1}.
\end{equation}

% \noindent\textbf{Localizer.}
We minimize the Binary Cross Entropy (BCE) loss between $\hat{\mathbf{M}}$ and $\mathbf{M}$.
\begin{equation}
\label{BCE_loss}
\begin{gathered}
{L}_{\emph{loc}}=\emph{BCE}(\mathbf{M},\hat{\mathbf{M}})\\
=-\frac{1}{H W}\sum(\mathbf{M}_{i,j}\log{\hat{\mathbf{M}}_{i,j}}+(1-{\mathbf{M}_{i,j}})\log{(1-\hat{\mathbf{M}}_{i,j}})).
\end{gathered}
\end{equation}

% \noindent\textbf{Discriminator.} 
% For adversarial training, the discriminators $\mathcal{D}_{A}$ and $\mathcal{D}_{B}$ need to respectively distinguish $\mathbf{X}$ and $\hat{\mathbf{I}}$ from $\mathbf{I}$. 
% The adversarial loss ${L}_{\emph{adv}}$ for the main pipeline is as follows.
% \begin{equation}
% {L}_{\emph{adv}}=\log{(1-\mathcal{D}_{A}(\hat{\mathbf{X}}))}+\log{(1-\mathcal{D}_{B}(\hat{\mathbf{I}}))}.
% \end{equation}
The loss for the two discriminators are respectively shown in Eq.(\ref{eqn_discriminator}), where we accept the most-common BCE loss.
\begin{equation}
\label{eqn_discriminator}
\begin{gathered}
{L}_{\mathcal{D}_{A}}=-\frac{1}{2}(\log{\mathcal{D}_{A}(\mathbf{I})}+\log{(1-\mathcal{D}_{A}(\mathbf{X}))}), \\
{L}_{\mathcal{D}_{B}}=-\frac{1}{2}(\log{\mathcal{D}_{A}(\mathbf{I})}+\log{(1-\mathcal{D}_{A}(\hat{\mathbf{I}}))}), \\
{L}_{\emph{GAN}}=-(\log{\mathcal{D}_{A}(\mathbf{X})}+\log{\mathcal{D}_{B}(\hat{\mathbf{I}}))}).
\end{gathered}
\end{equation}

% \noindent\textbf{feature alignment network.}
The performance consistency loss for the feature alignment network $\mathcal{F}$ is consist of four loss terms:
\begin{equation}
\label{eqn_student}
{L}_{\emph{cs}}=\lVert\hat{\mathbf{M}}^0-\hat{\mathbf{M}}^1\rVert_{1}+\lVert\hat{\mathbf{I}}^0-\hat{\mathbf{I}}^1\rVert_{1}+\lVert\hat{\mathbf{I}}^0-\mathbf{I}\rVert_{1}+\sum_{n\in [3,5]}\lVert{\phi}_{n}^{0}-{\phi}_{n}^{1}\rVert_{1},
\end{equation}
where $n\in [3,5]$ indicates the third to fifth layer of $\mathcal{F}$.

% \noindent\textbf{Total loss.}
The total loss for CLR-Net is based on the processed image provided by the student network $\mathbf{P}_{\emph{stu}}$.
\begin{equation}
\mathcal{L}={L}_{prt}+\alpha\cdot{L}_{\emph{rec}}+\beta\cdot{L}_{\emph{loc}}+\gamma\cdot{L}_{\emph{cs}}+\eta\cdot{L}_{\emph{GAN}}.
\end{equation}

\subsection{Network details}
\label{section_detail}
To implement $\mathcal{G}$, we use an invertible U-shaped network~\cite{xiao2020invertible} composed of three stacked downscaling
modules and three upsampling modules. 
% Each module contains a Haar downsampling or upsampling transformation layer and four double-sided affine coupling layers. 
% The wavelet-based blocks provide empirical biases to decompose the features into the lower-band and higher-band information, where the latter is essential for imperceptible image protection. 
We apply Spectral Normalization (SN) \cite{miyato2018spectral} in each block in that SN helps stabilize the training by restricting the Lipschitz constant within one.

For cropping simulation, 
% Table \ref{table_attack_summary} shows a summary of robustness against a variety of attacks and whether these attacks are introduced in $\mathcal{A}$. 
we \redmarker{freely} generate cropping masks $\mathbf{M}$ to crop the protected images. The \greenmarker{survival rate} is set as ${r}_{c}\in[0.5^2,1^2]$, where \greenmarker{$0.5^2$} stands for preserving only 25\% squared-size area of $\mathbf{X}$ while the rest are cropped. Afterward, we resize the cropped image and simulate common image post-processing attacks using classical differentiable lossy image operations \cite{zhu2018hidden}.
% including Gaussian blurring, median blurring, JPEG compression, Addition of White Gaussian Noise (AWGN) and rescaling. 
Lastly, we 
% use a differentiable image quantization method \cite{bengio2013estimating} to 
convert the floating-point values of $\mathbf{X}$ to 8-bit integer to save images on disk.

The feature alignment network $\mathcal{F}$ is sketched in Fig.~\ref{preprocessor}. 
It is built upon a six-layered fully-convolutional network.
Each \textit{Conv} block within the FCN shown in Fig.~\ref{preprocessor} consists of a \textit{Conv} layer, an ELU activation layer and a SN layer.

% FG-JPEG \greenmarker{needs to be} trained ahead of the pipeline with paired $\{\mathbf{I}$, $\mathbf{I}_{\emph{jpg}}\}$, where $\mathbf{I}_{\emph{jpg}}$ are the real-world JPEG images generated in advance. 
FG-JPEG consists of a generator, a controller and a predictor. Fig.~\ref{image_conponents} illustrates the network design. 
The predictor classifies the quality factor of a target image, which is implemented by ResNet-32 and trained to estimate the QF of real-world JPEG images $\mathbf{I}_{\emph{jpg}}$. 
We replace the leading \textit{Conv} layer with a vanilla \textit{Conv} layer, an SRM \textit{Conv} layer and a Bayar \textit{Conv} layer in parallel similar to \cite{wu2019mantra} to augment the image details.
% The introduction of SRM \textit{Conv} and Bayar \textit{Conv} is to augment details of the images, since JPEG compression generally alters the higher-frequent details and keeps intact the semantic information.
% For QF classification, the QF labels are set as $\emph{QF}\in\{10,30,50,70,90,100\}$ to ensure that the discrepancy among adjacent categories \redmarker{is} significant enough for classification. 
% The controller is to control the generation of simulated JPEG images, while the predictor is to \redmarker{estimate} whether the QF classification result of the output image can match that of the real-world JPEG image with the same QF.
% The \redmarker{estimated} quality factor can be changed with interactive selections to have a balance between artifacts removal and details preservation. 
The generator is built upon the popular four-leveled U-Net architecture, 
% In each level, there is a pair of a down-sampling block, an up-sampling block, and a channel-wise concatenation in between as a shortcut. 
% Traditional U-Net-based networks consist of an encoder part, a decoder part, a \textit{Conv} block linking the two parts and channel-wise concatenations in between the different levels as shortcuts. 
where we introduce an additional \textit{Conv} block to replace the straightforward concatenation in each level. 
% The implementation of the \textit{Conv} blocks is identical to those in the feature alignment network, and we use bilinear interpolation to conduct rescaling in the down-sampling and up-sampling blocks.
Furthermore, we let the outputs of the lower three levels of the additional \textit{Conv} blocks be conditioned on the output of the controller. 
The controller is a six-layered MLP, where the last three layers learn the mapping functions that output modulation parameter pairs ${a}, {b}$ that control the standard deviation and mean of the outputs.
% Specifically, the first three layers of the of MLP generate shared features, which are then split into three parts corresponding to the three scales in the generator. In the last layer of MLP, we learn different parameter pairs for different scales in the generator.
% The output of these additional Conv  blocks are
\begin{equation}
\mathbf{F}_{\emph{out},i}={a}_{i}\cdot\emph{Conv}(\mathbf{F}_{\emph{in},i})+{b}_{i},
\end{equation}
where $\mathbf{F}_{\emph{in}}^{i}$ and $\mathbf{F}_{\emph{out}}^{i}$ respectively denote the input and output features of the additional \textit{Conv} block at ${i}_{\emph{th}}$ level.
% The length of ${\gamma}_{i}, {\beta}_{i}$ is consistent with the number of channels of $\mathbf{F}_{ub,i-1}$. 
% We allow the network to give an imprecise prediction when QF lies in between two neighboring categories. 
The predictor then provides classification results $\hat{\emph{QF}}$ on $\hat{\mathbf{I}}_{\emph{jpg}}$, and the results should be close with those of $\mathbf{I}_{\emph{jpg}}$ with $\emph{QF}$. 
% When we apply FG-JPEG in the pipeline of CLR-Net, we consider that the difference of $\mathbf{I}$ and $\mathbf{X}$ lies mainly in the higher bands. Accordingly, we additionally introduce some Additive White Gaussian Noise (AWGN) on $\mathbf{I}$ and encourage the removal of higher-band details including AWGN during training FG-JPEG.
% We set $\delta_\emph{QF}=20$ among the neighboring categories where the discrepancy can be significant enough for classification. 
% We allow the network to give an imprecise \redmarker{estimation} when QF lies in between two neighboring categories.
Note that FG-JPEG should be trained ahead of CLR-Net, where we employ the Cross-Entropy (CE) loss.

\begin{equation}
\label{eqn_qf}
{L}_{\emph{QF}}=\emph{CE}(\mathbf{Q}_{o},\mathbf{Q}_{r})=-\sum_{c=1}^{6}y_{o,c}\log(p_{o,c}),
\end{equation}
where $y_{o,c}$ is the binary indicator if class label $c$ is the correct classification for observation $o$. $p_{o,c}$ is the \redmarker{estimated} probability observation $o$ is of class $c$.
% $\mathbf{Q}_{o}$ takes the \textit{argmax} of $o$ that maximizes $p$. 
The QF labels are set as $\emph{QF}\in\{10,30,50,70,90,100\}$
The JPEG generator and controller are jointly optimized by minimizing:
\begin{equation}
\label{eqn_jpeg}
{L}_{\emph{jpg}}=\lVert\mathbf{I}_{\emph{jpg}}-\hat{\mathbf{I}}_{\emph{jpg}}\rVert_{1}+\epsilon\cdot~\emph{CE}(\emph{QF},\hat{\emph{QF}}).
\end{equation}

The localizer $\mathcal{L}$ is built upon a three-layered lightweight U-Net that transforms $\hat{\mathbf{X}}_{\emph{atk}}$ into a one-dimensional feature. The feature is then flattened and fed into a four-layered MLP.
The output of the MLP is $\{x_0,y_0,x_1,y_1,S\}$, where $\{x_0,y_0,x_1,y_1\}$ represents the coordinates of the upper-left and lower-right corner of the rectangle-shaped $\hat{\mathbf{M}}$. $\mathbf{X}_{\emph{atk}}$ is \redmarker{detected} as \textit{cropped} if $S\geq~0.5$. 
We also feed $\mathcal{L}$ with non-cropped images $\mathbf{I}$ and encourage $S\approx~0$. 
% It lowers the false alarm rate of CLR-Net. 
% We resize and zero-pad $\hat{\mathbf{X}}_{\emph{atk}}$ according to the \redmarker{estimated} cropping mask. The intermediate image $\hat{\mathbf{X}}$ places the preserved contents in the correct location with the cropped-out contents left blank.
The two discriminators $\mathcal{D}_{A}$ and $\mathcal{D}_{B}$ are built upon the popular Patch-GAN \cite{isola2017image}.

\section{Experiments}
\label{sec:experiment}
% In this section, we provide comprehensive experiments to validate the effectiveness of CLR-Net. We begin with analyzing the performance of our method. Then we compare CLR-Net with some previous arts. Also, we validate the network design via ablation studies.
% \begin{figure*}
% 	\centering
% 	\includegraphics[width=1.0\textwidth]{figures/untrained_attacks.pdf}
% 	\caption{\textbf{Experimental results on untrained attacks.} (a) Brightness change \greenmarker{(+20)}, (b) automatic contrast enhancement \greenmarker{using CLAHE}, (c) hybrid attack (JPEG80+Resizing). We do not train the networks with brightness change, contrast enhancement since these attacks will shift the mean of the image, resulting a large variance from the ground-truth.}
% 	\label{image_generalization}
% \end{figure*}
% \begin{table}[!t]
% \footnotesize
% 	\renewcommand{\arraystretch}{1.}
% 	\caption{Average PSNR and SSIM between the original and protected images under different resolutions.}
% 	\label{table_different_resolution_protected}
% 	\begin{center}
% 	\begin{tabular}{c|cc|cc|cc}
% 		\hline
% 		 \multirow{2}{*}{Dataset} & \multicolumn{2}{c|}{${512}^2$} & \multicolumn{2}{c}{${256}^2$} & \multicolumn{2}{c}{${128}^2$} \\
% 		& PSNR & SSIM & PSNR & SSIM & PSNR & SSIM
% 		\\
%         \hline
%         COCO & 30.13 & 0.927 & 32.67 & 0.933 & 33.52 & 0.943\\
%         ImageNet & 30.01 & 0.927 & 33.01 & 0.937 & 33.33 & 0.940\\
%         % UCID & 29.52 & 0.920 & 32.15 & 0.920 & 33.01 & 0.935\\
%         Flicker & 31.03 & 0.932 & 33.13 & 0.940 & 33.78 & 0.951\\
% 		\hline
% 	\end{tabular}
% 	\end{center}
% \end{table}

\begin{table*}
\footnotesize
\renewcommand{\arraystretch}{1.}
    \caption{Average performance of cropping localization and image recovery over 1000 images from different datasets.}
    \label{table_comparison}
    \centering
	\begin{tabular}{c|c|c|ccc|ccc|cccc}
		\hline
		\multirow{2}{*}{\greenmarker{Survival rate}} &
		\multirow{2}{*}{Dataset} & \multirow{2}{*}{Index} & \multicolumn{3}{c|}{JPEG} & \multicolumn{3}{c|}{Scaling} &  \multirow{2}{*}{M-Blur} & \multirow{2}{*}{CE}&
		\multirow{2}{*}{Dropout}&\multirow{2}{*}{Identity} \\
		& & & QF90 & QF70 & QF50 & 150\% & 125\% & 75\% \\
        \hline
        \multirow{3}{*}{[$0.5^2$,$0.65^2$]}
        & \multirow{3}{*}{CelebA}
        & F1 & 0.955 & 0.940 & 0.921 & 0.956 & 0.933 & 0.947 & 0.935 & 0.827 & 0.864 & 0.980\\
        & & PSNR & 26.50 & 24.55 & 23.40 &  25.74 & 25.46 & 26.13 & 26.61 & 22.47 & 23.39 & 27.77\\
        & & SSIM & 0.790 & 0.753 & 0.719  & 0.797 & 0.770 & 0.814 & 0.816 & 0.701 & 0.710 & 0.829 \\
		\hline
		\multirow{3}{*}{[$0.65^2$,$0.8^2$]}& \multirow{3}{*}{COCO}
        & F1 & 0.968 & 0.970 & 0.951 &  0.963 & 0.970 & 0.956 & 0.941 & 0.912 & 0.920 & 0.986\\
        & & PSNR & 27.46 & 27.30 & 22.20  & 26.34 & 24.94 & 25.75 & 26.34 & 23.04 & 25.46 & 30.18\\
        & & SSIM & 0.866 & 0.854 & 0.707 & 0.801 & 0.767 & 0.769 & 0.755 & 0.731 & 0.789 & 0.891 \\
		\hline
        \multirow{3}{*}{[$0.8^2$,$1.0^2$]} & \multirow{3}{*}{Places} 
        & F1 & 0.982 & 0.979 & 0.965 & 0.970 & 0.958 & 0.969 & 0.947 & 0.901 & 0.885 & 0.990\\
        & & PSNR & 29.70 & 27.55 & 25.92 & 27.69 & 27.62 & 29.29 & 26.80 & 26.62 & 22.09 & 31.07\\
        & & SSIM & 0.844 & 0.819  & 0.774 & 0.853 & 0.846 & 0.849 & 0.811 & 0.772 & 0.740 & 0.898\\
		\hline
% 		\multirow{4}{*}{\rotatebox{90}{ImageNet}} & BCE & 0.022 & 0.096 & 0.106 & 0.148 & 0.074 & 0.121 & 0.027 & 0.082 & 0.015\\
%         & F1 & 26.11 & 23.85 & 18.71 & 23.87 & 25.10 & 24.52 & 26.85 & 24.73 & 27.11\\
%         & PSNR & 29.33 & 27.43 & 25.98 & 25.23 & 26.61 & 25.78 & 28.66 & 25.94 & 29.60\\
%         & SSIM & 0.862 & 0.854  & 0.803 & 0.838 & 0.870 & 0.829 & 0.863 & 0.821 & 0.865 \\
% 		\hline
	\end{tabular}
\end{table*}

% \subsection{Experimental settings}
% \label{section_experimental_setting}
% \noindent\textbf{Data preparation. }

During training, the original images $\mathbf{I}$ in our scheme are from a mixture of multiple popular image datasets, namely,  COCO~\cite{lin2014microsoft}, 
CelebA~\cite{liu2018large},
% Flicker~\cite{flickr1024}, 
% UCID~\cite{schaefer2003ucid},  
% ImageNet~\cite{johnson2016perceptual},
Places~\cite{zhou2017places} and ImageNet.
% , Paris Street View~\cite{PASSRnet} and DIV2K~\cite{agustsson2017ntire}.
% Fig.~\ref{image_main_results} showcases four images that are arbitrarily sampled from COCO, Places, CelebA and Paris Street View.
We arbitrarily select around 10000 images from the training set of the above datasets. We build the test set in the same way with 1000 images from each.
% \greenmarker{We do not build the validation set and train the networks for 20 epochs.}
% Unlike some previous schemes~\cite{ying2021no,zhu2018hidden} that train individual networks for different attacks or datasets, we train a universal model for all possible image attacks, but we train different models for varied resolutions considering that convolutions are not scale-agnostic.
% Thus, CLR-Net can be applied on images with arbitrary resolution using the most suitable model. 
In the following experiments, we resize the images into size $256\times256$, and the results under other typical resolutions are close. 
% The scheme is tested with human-participated real-world attacks. The protected images are saved in 8-bit PNG format and the attacked images are randomly saved in JPEG, BMP or PNG format. 
The hyper-parameters are set as $\alpha=1.5, \beta=0.1, \gamma=0.05, \epsilon=0.1$ and $\eta=0.01$. 
% According to our experiments, these hyper-parameters provide the best overall performance. 
% \greenmarker{To better control the quality of the protected image, we additionally regulate that if the average PSNR between the original and protected image is lower than an expected threshold, e.g., $32.5$dB, we let $\alpha=4$, which encourages lower magnitude of embedding.}
% The computational complexity of CLR-Net is comparatively low. The networks are in sum composed of 7.84 million trainable parameters. \greenmarker{We train CLR-Net on two Nvidia RTX 3090 GPUs and} in each second CLR-Net can infer 8.63 images during testing. 
The batch size of CLR-Net during training is set as 4.
We use Adam optimizer with the default parameters. The learning rate is $1\times10^{-4}$ with \redmarker{cosine annealing} decay.

\subsection{Qualitative and quantitative analysis}
\noindent\textbf{Quality of the protected images.} 
Fig.~\ref{image_main_results} showcases four experimental results of CLR-Net on cropping localization and image recovery.
We can observe that the quality of $\mathbf{X}$ is satisfactory with the difference between $\mathbf{X}$ and $\mathbf{I}$ almost imperceptible. 
% A closer look at the augmented difference indicates that the information of the entire image is encoded in the higher bands \greenmarker{and mainly within the image center. 
% Also, to combat JPEG compression, we see that the embedded information are generally encoded into the chessboard style.}
We have conducted quantitative embedding experiments on different datasets. 
The averaged PSNR and SSIM between $\mathbf{I}$ and $\mathbf{X}$ is respectively 32.67dB/ 0.933 on COCO, 33.01dB/ 0.937 on ImageNet, and 33.13dB/ 0.940 on Flicker. 
% and the results in Table~\ref{table_different_resolution_protected} show that more perturbation is required for textured image datasets and larger images. 
% For example, for ordinary datasets like COCO, the average PSNR and SSIM are respectively 32.67dB \greenmarker{and 0.933}. UCID contains way more textured images, and the PSNR and SSIM drop to 32.15dB and 0.920.
% \greenmarker{Also, on COCO, if the images are sized $256\times~256$, the averaged PSNR between the original and protected images will be close to 2.5dB higher than that of $512\times~512$-sized images.}

\noindent\textbf{Performance \greenmarker{of cropping localization and recovery.}} 
Fig.~\ref{image_main_results} further shows the results of cropping localization and image recovery.
% The tests are made by the volunteers with different \greenmarker{survival rate} and image processing attacks. 
% \greenmarker{Because the settings of the attacks during training simulate most commonly used image post-processing attacks, we employ the same settings during testing. For example, we arbitrarily set the kernel size as three or five when we apply the Gaussian blurring attack.}
We can observe that the original images are recovered even with the median blurring attack and a medium-sized cropping mask.
The average PSNR between $\mathbf{I}$ and $\hat{\mathbf{I}}$ in these cases is 26.73dB with SSIM 0.843. The cropping localization is accurate, with the average F1 score close to 0.95. 
% Though it is not possible to predict a perfect $\hat{\mathbf{M}}$, CLR-Net learns to adapt to the trivial drift. 
% Besides, though some details of the cropped-out contents are lost, the fidelity of the recovered contents is high in that CLR-Net does not produce them by hallucination.
We have conducted more experiments over the test set under different \greenmarker{survival rate} and benign attacks. The average performances are reported in Table~\ref{table_comparison}. 
% As a clarification, ``Scaling 150\%" is that after cropping, the attacker upscales the cropped image where the scaling factor is 1.5. 
% We see that heavier attacks or larger \greenmarker{survival rates} will lower the performance of both cropping localization and recovery.
For image recovery, the average SSIMs are generally above 0.8, which can provide satisfying and trustworthy recovery results. 
% The F1 scores are generally above 0.9, which indicates cropping localization is easier than image recovery.
% \noindent\textbf{\greenmarker{Performance} generalization on untrained attacks.}
% We further test CLR-Net on several untrained attacks, namely, brightness change, contrast enhancement and hybrid attack.
% \greenmarker{For brightness change, each pixel within the image is subtracted with a same digit arbitrarily ranging from $[-20.20]$ where the resulting overflow/underflow pixels will be clamped.
% For contrast enhancement, we employ the automatic Contrast Limited Adaptive Histogram Equalization (CLAHE)~\cite{CLAHE} implemented by the OpenCV library.
% For drop-out, the dropping-out rate is arbitrarily selected from $[0,0.3]$.
% } 
% We see that the results are also satisfactory, where the cropped part can be effectively localized.
% Interestingly, the recovered contents tend to be more close to the ground truth rather than the enhanced fragment. As a result, we can observe some color inconsistency between the cropped part and the recovered contents in the upper two examples. 
% It further proves that CLR-Net focuses on fidelity rather than feasibility during image recovery. 
Table~\ref{table_comparison} also reports quantitative results on untrained attacks, where the performance gaps between trained and untrained attacks are acceptable.

\subsection{Comparisons}
\noindent\textbf{Image recovery from the cropped version}. 
There is no previous work that recovers the entire image from its cropped version. 
In Table 1, we compare our method with \cite{ying2021image} by extending it from manipulated content recovery to image cropping recovery.
% We trained the models from scratch for cropping recovery alone. 
We find that though the method can somehow localize the cropped region well, the performance of \cite{ying2021image} concerning cropping recovery has a noticeable gap compared to ours.
% Details of our implementation of [B] for cropping recovery is introduced in Section 3.2 of the rebuttal.
Besides, we further compare the method with Stable Diffusion-Infinity~\footnote{https://huggingface.co/spaces/lnyan/stablediffusion-infinity}, which is typically designed for outpainting. 
We find that though the network can produce extra-high-quality results with astonishing details and imagination, yet the outpainted contents cannot ensure fidelity with reference to the original ones.
In contrast, CLR-Net pursues the ``exactness" of cropped content recovery, which emphasizes more on fidelity rather than ``feasibility".

% We build a baseline that trains two independent networks for a solution of Eq.~(\ref{equation_problem_statement}), where image protection is not involved. We implement the recovery network \redmarker{using} a state-of-the-art and open-source image outpainting network~\cite{van2019image}. The architecture of the localizer and the experimental settings are identical to those of CLR-Net.
% \begin{figure}[!t]
% 	\centering
% 	\includegraphics[width=0.48\textwidth]{figures/comparison_benchmark.pdf}
% 	\caption{\textbf{Comparison of CLR-Net with the baseline on two JPEG images}. CLR-Net can preserve much better image fidelity where the recovered contents are not hallucinated.}
% 	\label{image_compare_outpainting}
% \end{figure}
\begin{figure}
  \includegraphics[width=0.5\textwidth]{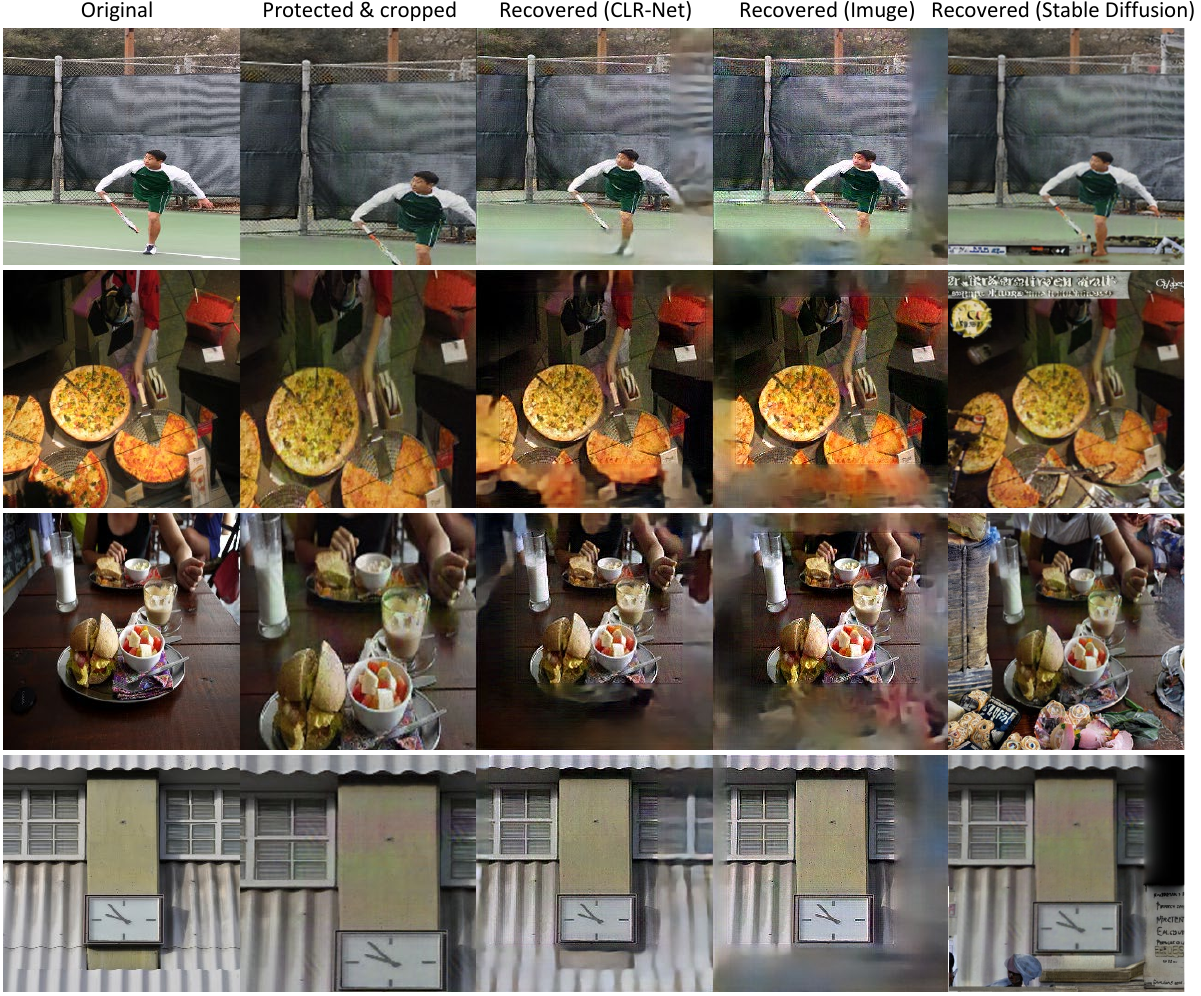}
  \caption{
  \textbf{Comparison of cropped image recovery with \cite{ying2021image} and outpainting, a related technology.} Owing to more sophisticated and targeted design of network bias and FG-JPEG, our proposed method surpasses \cite{ying2021image}  noticeably.}
  \label{rebuttal_comparison}
\end{figure}
\setlength{\tabcolsep}{1.05mm}{
\begin{table}[!t]
\footnotesize
    \caption{\textbf{Quantitative performance comparison between our method (CLR-Net, 1000 images), Imuge~\cite{ying2021image} (1000 images) and outpainting (Stable Diffusion, 100 images).} For \cite{ying2021image}, we control that the PSNR before and after image protection should be around 33dB for fair comparison.}
    	\label{table_comparison_hybrid}
    \centering
	\begin{tabular}{c|cc|cc|cc|cc}
		\hline
		\multirow{2}{*}{Method} & 
		 \multicolumn{2}{c|}{NoAtk} &
		 \multicolumn{2}{c|}{JPEG70} &
		 \multicolumn{2}{c|}{Scaling} &
		 \multicolumn{2}{c}{M-Blur}\\
        \cline{2-9}
        & F1 & PSNR 
        & F1 & PSNR 
        & F1 & PSNR
        & F1 & PSNR\\
        \hline
        Imuge [B]
        & .944 & 26.55 & .909 & 23.97 & .922 & 25.48 & .885 & 24.15 \\
        Stable Diffusion
        & - & 17.42 & - & 17.63 & - & - & - & - \\
        Ours (CLR-Net)
        & .986 & 30.18 & .970 & 27.30 & .963 & 26.34 & .947 & 26.80 \\
		\hline
  
	\end{tabular}
\end{table}
}

\noindent\textbf{Cropping identification and localization}.
% \greenmarker{Image cropping localization is historically less investigated because traces left by cropping is very subtle and fragile~\cite{van2020dissecting}.} 
% So far, the state of the art in passive image cropping prediction accuracy is held by Van et al.~\cite{van2020dissecting}, whose 
The overall accuracy of Van et al.~\cite{van2020dissecting} is 86\% on uncompressed images. Whenever the input images are compressed or tainted by other digital attacks, the scheme can no longer predict the cropping mask. 
The cropping prediction accuracy of CLR-Net is 95.41\% that further leads by a large margin. 
Note that in the result we include the false alarm rate where we should not \redmarker{predict} \textit{positive} on non-cropped images. 
% The reason is that CLR-Net does not have restrictions on the quality or format of the targeted image, unlike Van et al.~\cite{van2020dissecting} where the statistical clue of lens artifacts are prone to compression.
% , the scheme is not applicable on the majority of natural images.
% According to Table~\ref{table_comparison}, the F1 scores of CLR-Net under JPEG ($\emph{QF}=70$) is within $[0.94,0.98]$, which is also higher than 0.74 reported in~\cite{ying2021no}. 
% We owe the high accuracy of image cropping detection of CLR-Net to robust watermarking. 

% ------------------------------------------
% ------------move to supplement------------
% ------------------------------------------
% The quality factor is respectively 10 in the first row and 80 in the second row. FG-JPEG contains less artifacts and compared with Diff-JPEG, and the outputs are more close to the ground truth compressed image.

% ------------------------------------------
% ------------move end------------
% ------------------------------------------

\subsection{Ablation studies}
\label{section_ablation}

\begin{figure}[!t]
	\centering
	\includegraphics[width=0.48\textwidth]{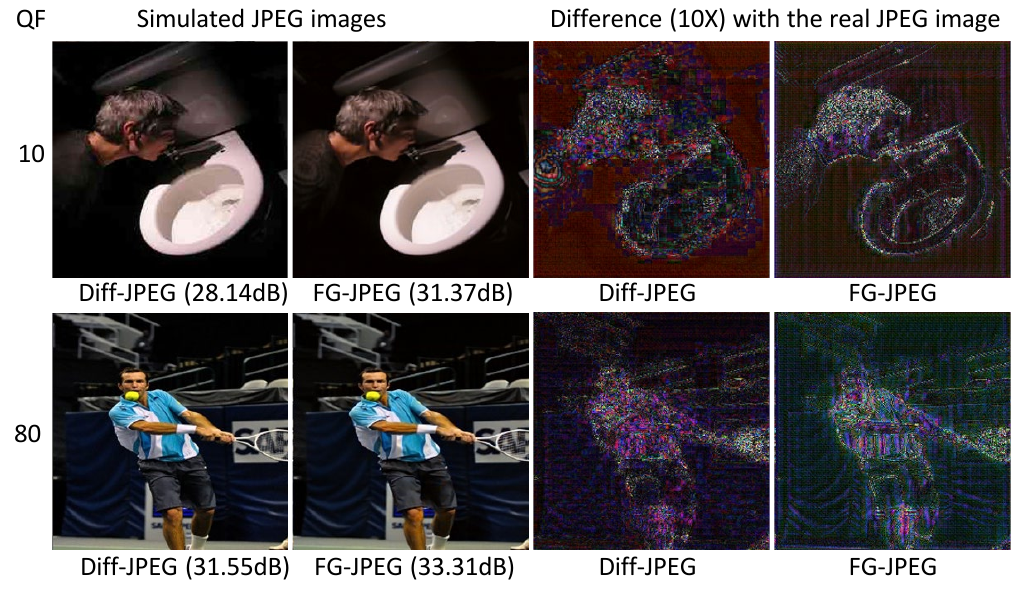}
	\caption{\textbf{Comparison on the fidelity of the simulated JPEG images} between FG-JPEG and Diff-JPEG~\cite{shin2017jpeg}. }
	\label{image_jpeg_simulation}
\end{figure}

\noindent\textbf{Influence of feature alignment and FG-JPEG.}
We see a promising performance gain by the feature alignment operation. \redmarker{It proves} that the network \greenmarker{can} help unify $\mathbf{I}_{\emph{atk}}$ made by different benign attacks, and therefore allow the generator to work on a \greenmarker{more clustered} transformed domain.
% Besides, using the popular Diff-JPEG~\cite{shin2017jpeg} as the JPEG simulator to train our network also gives inferior performances.
From Fig.~\ref{image_jpeg_simulation} and Table~\ref{table_jpeg_compare}, the PSNR results suggest that our generative method gives closer results to the real-world JPEG images. 
The averaged QF classification accuracy of Diff-JPEG~\cite{shin2017jpeg} is 72.48\%, while that of FG-JPEG is 93.17\%. 
It indicates that more knowledge about JPEG compression is learned by FG-JPEG with the joint guidance of the generative model and the QF predictor. 
Besides, the chess-board artifact of JPEG compression can also be found in the generated JPEG images.

\noindent\textbf{Influence of INN architecture.}
A typical alternative is to model image protection and image recovery independently using the traditional  \textit{encoder-decoder} networks. 
\greenmarker{Thus, we implement the image protection network and the image recovery network using two simple U-Nets. For a fair comparison, we regulate that the two networks contain the same number of parameters in sum with CLR-Net.}
From Table~\ref{table_ablation}, we observe that INN-based CLR-Net provides better performance.

% \begin{figure}[!t]
% 	\centering
% 	\includegraphics[width=0.48\textwidth]{figures/fig9.pdf}
% 	\caption{Ablation studies of CLR-Net via observing the quality of the recovered images.}
% 	\label{image_ablation}
% \end{figure}
\begin{table}[!t]
\footnotesize
	\renewcommand{\arraystretch}{1.}
	\caption{\greenmarker{\textbf{Comparison of different JPEG simulators} by measuring the average PSNR between the ground-truth and simulated JPEG images (QF=70).}}
	\label{table_jpeg_compare}
	\begin{center}
	\begin{tabular}{c|c|c|c|c}
		\hline
		 & FG-JPEG & Diff-JPEG & JPEG-Mask & JPEG-SS
		\\
        \hline
        PSNR & 32.55 & 31.42 & 28.49 & 30.77 \\
        Accuracy & 93.17\% & 72.48\% & 64.55\% & 73.42\% \\
		\hline
	\end{tabular}
	\end{center}
\end{table}
\begin{table}[!t]
\footnotesize
	\renewcommand{\arraystretch}{1.}
\caption{\textbf{Averaged results of ablation studies} on COCO. The tests are done under JPEG attack (QF=70) and $r_{c}=0.7^2$.}
    	\label{table_ablation}
    \begin{center}
    \begin{tabular}{c|c|c|c}
    \hline
    		
    		 Setting & F1 & PSNR & SSIM\\
    		% \hline
    		% w/o image protection ($\mathbf{I}=\mathbf{X}$)  & 0.435 & 18.13  & 0.585 \\
      \hline
    		% \hline
    		% w/o multi-staged training and synchronization & 0.268 & 11.35  & 0.244 & 0.163 & 7.49 & 0.228 \\
    		% \hline
    		% w/o gradient stabilization (do not prevent gradient explosion) & 0.624 & 16.73  & 0.527 & 0.255 & 10.09 & 0.297\\
    		% \hline
    		w/o feature alignment ($\hat{\mathbf{X}}_{\emph{atk}}=\mathbf{X}_{\emph{atk}}$) & 0.958 & 24.12  & 0.736 \\
    		% \hline
    		% w/o tamper-based data augmentation & 0.802 & 23.61  & 0.701 & 0.844 & 24.65 & 0.762\\
			\hline
    		separate $\mathbf{G}$ using two U-Nets  & 0.933 & 21.54 & 0.713 \\
    		% \hline
    		% $\mathbf{A}$ using JPEG-Mask~\cite{zhu2018hidden} & 0.917 & 22.47  & 0.678 & 0.923 & 23.07 & 0.699\\
    % 		\hline
    % % 		$\mathbf{A}$ using Diff-JPEG~\cite{shin2017jpeg} & 0.962 & 23.78  & 0.755\\
    		\hline
    		$\mathbf{A}$ using Diff-JPEG~\cite{shin2017jpeg} & 0.958 & 25.95  & 0.814\\
    		\hline
    		CLR-Net (Full implementation) & $\textbf{0.965}$ & $\textbf{27.02}$  & $\textbf{0.822}$ \\
    		\hline
    	\end{tabular}
    	\end{center}
\end{table}

\section{Conclusion}
\label{section_conclusion}
This paper proposes an image protection scheme for image cropping localization and recovery.
We formulate a model of the problem upon an \redmarker{INN}-based anti-crop generator as well as a crop localizer.
% Besides, we propose two plug-in-and-play gadgets in our network to enhance real-world robustness. 
% First, a novel Fine-Grained generative JPEG simulator (FG-JPEG) is used to flexibly mimic the characteristics of JPEG compression. Second, a Siamese network is proposed for image pre-processing on the attacked images where we minimize the performance gap against two different attacks in each iteration. 
Comprehensive experiments prove the effectiveness of CLR-Net.
% CLR-NET can resist typical image attacks, where the novel FG-JPEG as well as the image pre-processor help promote real-world robustness. 
% In future works, the performance of CLR-Net might be improved with the introduction of long-range attention-based mechanisms.

\bibliographystyle{IEEEbib}
\bibliography{icme2023template}

\begin{thebibliography}{10}

\bibitem{fanfani2020vision}
Marco Fanfani, Massimo Iuliani, Fabio Bellavia, Carlo Colombo, and Alessandro
  Piva,
\newblock ``A vision-based fully automated approach to robust image cropping
  detection,''
\newblock {\em Signal Processing: Image Communication}, vol. 80, pp. 115629,
  2020.

\bibitem{yerushalmy2011digital}
Ido Yerushalmy and Hagit Hel-Or,
\newblock ``Digital image forgery detection based on lens and sensor
  aberration,''
\newblock {\em International Journal of Computer Vision}, vol. 92, no. 1, pp.
  71--91, 2011.

\bibitem{li2009passive}
Weihai Li, Yuan Yuan, and Nenghai Yu,
\newblock ``Passive detection of doctored jpeg image via block artifact grid
  extraction,''
\newblock {\em Signal Processing}, vol. 89, no. 9, pp. 1821--1829, 2009.

\bibitem{van2020dissecting}
Basile Van~Hoorick and Carl Vondrick,
\newblock ``Dissecting image crops,''
\newblock in {\em Proceedings of the IEEE/CVF International Conference on
  Computer Vision}, 2021, pp. 9741--9750.

\bibitem{zhu2018hidden}
Jiren Zhu, Russell Kaplan, Justin Johnson, and Li~Fei-Fei,
\newblock ``Hidden: Hiding data with deep networks,''
\newblock in {\em Proceedings of the European Conference on Computer Vision
  (ECCV)}, 2018, pp. 657--672.

\bibitem{ying2021image}
Qichao Ying, Zhenxing Qian, Hang Zhou, Haisheng Xu, Xinpeng Zhang, and Siyi Li,
\newblock ``From image to imuge: Immunized image generation,''
\newblock in {\em Proceedings of the 29th ACM international conference on
  Multimedia}, 2021, pp. 1--9.

\bibitem{shin2017jpeg}
Richard Shin and Dawn Song,
\newblock ``Jpeg-resistant adversarial images,''
\newblock in {\em NIPS 2017 Workshop on Machine Learning and Computer
  Security}, 2017, vol.~1.

\bibitem{xiao2020invertible}
Mingqing Xiao, Shuxin Zheng, Chang Liu, Yaolong Wang, Di~He, Guolin Ke, Jiang
  Bian, Zhouchen Lin, and Tie-Yan Liu,
\newblock ``Invertible image rescaling,''
\newblock in {\em European Conference on Computer Vision}. Springer, 2020, pp.
  126--144.

\bibitem{miyato2018spectral}
Takeru Miyato, Toshiki Kataoka, Masanori Koyama, and Yuichi Yoshida,
\newblock ``Spectral normalization for generative adversarial networks,''
\newblock {\em arXiv preprint arXiv:1802.05957}, 2018.

\bibitem{wu2019mantra}
Yue Wu, Wael AbdAlmageed, and Premkumar Natarajan,
\newblock ``Mantra-net: Manipulation tracing network for detection and
  localization of image forgeries with anomalous features,''
\newblock in {\em Proceedings of the IEEE/CVF Conference on Computer Vision and
  Pattern Recognition}, 2019, pp. 9543--9552.

\bibitem{isola2017image}
Phillip Isola, Jun-Yan Zhu, Tinghui Zhou, and Alexei~A Efros,
\newblock ``Image-to-image translation with conditional adversarial networks,''
\newblock in {\em Proceedings of the IEEE Conference on Computer Vision and
  Pattern Recognition}, 2017, pp. 1125--1134.

\bibitem{lin2014microsoft}
Tsung-Yi Lin, Michael Maire, Serge Belongie, James Hays, Pietro Perona, Deva
  Ramanan, Piotr Doll{\'a}r, and C~Lawrence Zitnick,
\newblock ``Microsoft coco: Common objects in context,''
\newblock in {\em European Conference on Computer Vision}. Springer, 2014, pp.
  740--755.

\bibitem{liu2018large}
Ziwei Liu, Ping Luo, Xiaogang Wang, and Xiaoou Tang,
\newblock ``Large-scale celebfaces attributes (celeba) dataset,''
\newblock {\em Retrieved August}, vol. 15, no. 2018, pp. 11, 2018.

\bibitem{zhou2017places}
Bolei Zhou, Agata Lapedriza, Aditya Khosla, Aude Oliva, and Antonio Torralba,
\newblock ``Places: A 10 million image database for scene recognition,''
\newblock {\em IEEE transactions on pattern analysis and machine intelligence},
  vol. 40, no. 6, pp. 1452--1464, 2017.

\end{thebibliography}

\end{document}